%% file: paper.tex
\documentclass[twoside,leqno,twocolumn]{article}
\pdfoutput=1
\usepackage[letterpaper]{geometry}
\usepackage{ltexpprt}
\usepackage{amsmath,amssymb,amsfonts}
\usepackage{algorithmic}
\usepackage[dvipdfm]{graphicx} 
\usepackage{bmpsize}
\usepackage{textcomp}
\usepackage{xcolor}
\usepackage[stretch=20,shrink=70]{microtype}
\usepackage{float}
\usepackage{booktabs}
\usepackage{multirow}
\usepackage{enumitem}
\usepackage{tikz}
\usepackage{mdframed}
\usepackage{balance}
\usepackage{wasysym}
\usepackage{graphicx}
\usepackage{subcaption}
\usepackage{url}
\usepackage{hyperref}
\definecolor{royalblue}{rgb}{0.25, 0.41, 0.88}
\definecolor{darkred}{rgb}{0.55, 0.0, 0.0}
\hypersetup{
    colorlinks = true,
    citecolor = royalblue,
    linkcolor = darkred,
    urlcolor = darkred,
}

\def\BibTeX{{\rm B\kern-.05em{\sc i\kern-.025em b}\kern-.08em
    T\kern-.1667em\lower.7ex\hbox{E}\kern-.125emX}}
\input{definitions}

\begin{document}


\title{\Large Evidential Cyber Threat Hunting}

\author{\normalsize Frederico Araujo
\quad Dhilung Kirat
\quad Xiaokui Shu
\quad Teryl Taylor
\quad Jiyong Jang\\
\normalsize IBM Research, Yorktown Heights, NY, USA\vspace{-12pt}}

\date{\vspace{-16pt}}


\fancyfoot[R]{\scriptsize{Copyright \textcopyright\ 2021 by SIAM\\
Unauthorized reproduction of this article is prohibited}}




\pagenumbering{arabic}
\setcounter{page}{1}

\maketitle

\parskip=0pt
\frenchspacing

\advance\baselineskip-.85pt plus.2pt minus.5pt

\input{sections/abstract}

\input{sections/introduction}

\input{sections/hunting-model}

\input{sections/case-study}

\balance

\input{sections/implementation}

\input{sections/protect-measures}

\input{sections/conclusion}

\section*{Acknowledgments}
The research reported herein was developed with funding from the Defense Advanced Research Projects Agency (DARPA) under the U.S.~ACC-APG/DARPA award W912CG-19-C-0003. The views, opinions and/or findings expressed are those of the author and should not be interpreted as representing the official views or policies of the Department of Defense or the U.S. Government. The U.S. Government is authorized to reproduce and distribute reprints for Government purposes notwithstanding any copyright notation hereon. Approved for Public Release, Distribution Unlimited.

{
 \footnotesize
 \global\let\oldthebib\thebibliography
 \renewcommand\thebibliography[1]{\oldthebib{#1}\itemsep0pt plus.5pt minus0pt\relax\itemindent-4pt}%
 \bibliographystyle{abbrv}%
 \bibliography{bibliography}
}

\end{document}

%% file: definitions.tex
\newcommand\contrib[1]{}
\newcommand{\showOwners}[1]{}
\newcommand{\kb}{K}
\newcommand{\rawdata}{\Sigma}
\newcommand{\intel}{\mathcal{I}}
\newcommand{\actions}{\mathcal{A}}
\newcommand{\hyps}{\mathcal{H}}
\newcommand{\dhyps}{\mathcal{H_{D}}}
\newcommand{\thyps}{\mathcal{H_{T}}}
\newcommand{\detector}{\delta}
\newcommand{\detectors}{\Delta}
\newcommand{\hunter}{\phi}
\newcommand{\hunters}{\Phi}
\newcommand{\manifold}{\psi}
\newcommand{\manifolds}{\Psi}
\newcommand{\verifier}{\gamma}
\newcommand{\verifiers}{\Gamma}

\newcommand\amper{{\sf\small\&}}

\newcommand{\fig}{\textcolor{darkred}{Figure~}}
\newcommand{\tab}{\textcolor{darkred}{Table~}}
\newcommand{\s}{\textcolor{darkred}{\S}}

\newcommand*\circled[2][.75pt]{\tikz[baseline=(char.base)]{
            \node[shape=circle,draw,inner sep=#1] (char) {\footnotesize#2};}}
            
\newcommand{\inlinedsection}[2][6pt]{\vspace{#1}\noindent\small\textbf{#2}}

\newcommand{\inlinedsectionitblock}[2][6pt]{\vspace{#1}\noindent\textit{#2}}
\newcommand{\inlinedsectionit}[2][6pt]{\vspace{#1}\noindent\textit{#2}\vspace{#1}}

\mdfdefinestyle{frame}{%
topline=false,rightline=false,leftline=false,
linewidth=0.5pt, 
skipabove=0pt,
skipbelow=0pt,
leftmargin=0pt,
rightmargin=0pt,
innerleftmargin=0pt,
innerrightmargin=0pt,
innertopmargin=-5pt,
innerbottommargin=5pt}

%% file: sections/abstract.tex
\begin{abstract}  \small\baselineskip=9pt
{\it
A formal cyber reasoning framework for automating the threat hunting process is described. The new cyber reasoning methodology introduces an operational semantics that operates over three subspaces---knowledge, hypothesis, and action---to enable human-machine co-creation of threat hypotheses and protective recommendations. An implementation of this framework shows that the approach is practical and can be used to generalize evidence-based multi-criteria threat investigations.
}



\end{abstract}

%% file: sections/introduction.tex
\section{Introduction}
\label{sec:intro}

Traditional attack detection and defense mechanisms often operate as separate, loosely-connected activities in threat hunting and mitigation contexts. When a suspicious event is detected in the monitoring data, the conventional course of action is to generate an intrusion alert in order to notify a security analyst about a potential threat. In response to the incident, the analyst correlates the alert to different sources of intelligence (both internal and external to the organization) to determine an appropriate mitigation strategy to counter the threat. The outcome is the execution of one or more defensive actions, such as blacklisting a remote DNS domain, patching a software vulnerability, or isolating potentially affected assets for further investigation.

\begin{figure}[t]
	\centering
	\begin{subfigure}[h]{0.185\textwidth}
		\includegraphics[width=\textwidth]{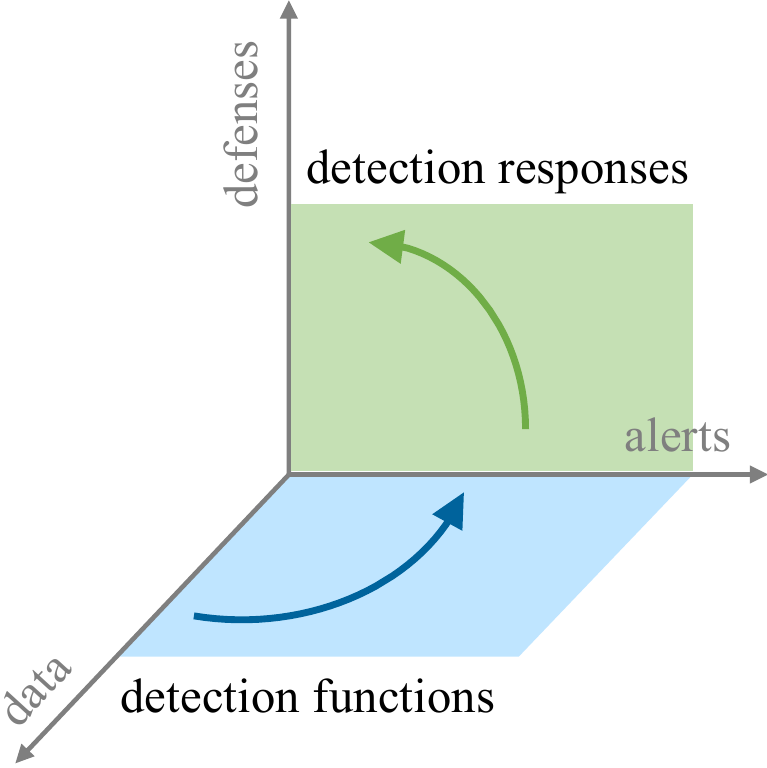}
		\caption{}
		\label{fig:manifold:traditional}
	\end{subfigure}
	~
	\begin{subfigure}[h]{0.275\textwidth}
		\includegraphics[width=\textwidth]{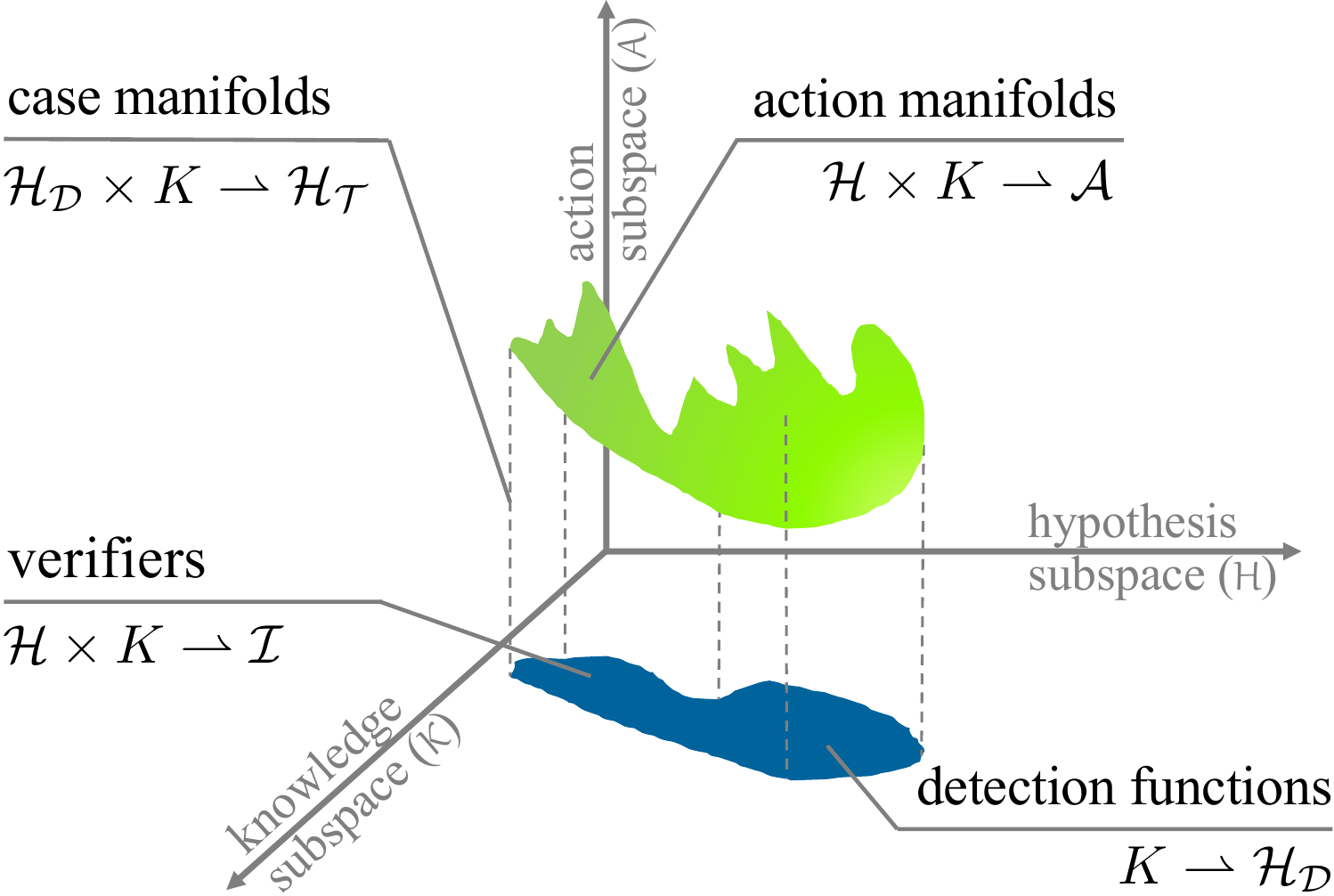}
		\caption{}
		\label{fig:manifold:HDMF}
	\end{subfigure}
	\caption{Illustration of (a)~traditional first-order functions in threat detection and response, and (b)~high-dimensional decision manifolds in cyber threat hunting.}
\label{fig:manifold}
\vspace{-10pt}
\end{figure}

This semi-manual process is onerous and requires that cyber combatants connect 
spatially and temporally scattered security alerts in order to comprehend threat scenarios, assess the impact of an attack, and defend against it. \fig\ref{fig:manifold:traditional} illustrates this concept. In this model, threat detectors are \emph{partial functions} mapping collected \emph{data} (e.g., network packets, system traces, logs) to security \emph{alerts}, and detection responses are denoted by the hyperplane relating alerts to \emph{defenses}---the protective measures available to the defendant. Such a reactive approach is insufficient for effective threat hunting because (a)~it fails to capture the high-dimensional features of advanced attacks (local, global, and offense contexts), yielding too many false positives, (b)~it imposes a \emph{cognitive overload}~\cite{Kirsh2000} on security analysts,
and (c)~it misses the larger attack campaign, because defensive actions respond to single, out-of-context events, therefore promoting a ``whack-a-mole'' threat detection methodology.

Recent automated approaches employ provenance tracking to discover causality between alerts~\cite{hassan2020tactical}, prune search trees~\cite{hassan2019nodoze, han2020unicorn} and link low-level data to domain knowledge~\cite{milajerdi2019holmes, hossain2017sleuth}. However, while useful in combating threat alert fatigue, 
these approaches lack support for iterative reasoning, and do not offer a framework for integrating data from various telemetry and intelligence sources, and making explainable decisions from such data.

To better characterize the dynamics of advanced cyber threats, 
this paper 
formalizes a cyber reasoning framework 
based on \emph{multi-functors} that define structure-preserving mappings (a.k.a.~morphisms) on threat knowledge, hypothesis, and action subspaces. Unlike the \emph{first-order} threat detection and response capabilities of traditional cyber security models, such decision \emph{manifolds}, depicted in \fig\ref{fig:manifold:HDMF}, connect cyber-hunting operations in a high-dimensional space---where each subspace comprises many dimensions---thus coping with the cognitive overload problem (by enriching sensor data with threat information that can be used to automatically reject or accept threat hypotheses) and adversarial targeting (by embedding attack models into the decision space).

Our framework
introduces three high-dimensional subspaces, which shape decision boundaries and augment the threat hunting process with contextual information, evidence-based reasoning, and a defense capability model. 

\inlinedsectionitblock{Knowledge Subspace} defines the scope of what is \emph{known} about the \emph{internal} and \emph{external} environments. Internal knowledge includes monitoring events (e.g., firewall and system logs, network packets and flows, system traces), network and assets information (e.g., sensors, services, version, configuration), and internal threat information (e.g., vulnerability assessment reports, verified threats). External knowledge includes threat intelligence~\cite{Li2019} (e.g., security feeds, threat reports, malicious domains) and security analytics information (e.g., security knowledge graphs, file reputation and analysis services). 

\label{sec:intro:subspaces}
\inlinedsectionitblock{Hypothesis Subspace} defines the domains of \emph{detection hypotheses} and \emph{threat hypotheses}. A detection hypothesis defines an attack proposition (e.g., ``host$_\textnormal{A}$ is vulnerable to reflected server XSS'', ``host$_\textnormal{B}$ and client$_\textnormal{X}$ beacon every two hours''),
which can be substantiated by the type of detection analysis employed---encoding a notion of detection precision, or confidence level in the hypothesis. A threat hypothesis embeds the higher-level reasoning performed by threat hunters, serving as cognitive glue and trigger for the next step of the hunting process. Examples of threat hypotheses include propositions about threat type (e.g., ``downloaded binary in client$_\textnormal{Y}$ is trojan$_\textnormal{Z}$'', ``beaconing behavior between host$_\textnormal{B}$ and client$_\textnormal{X}$ is part of a C\amper{}C protocol'')  and impact assessment (e.g., ``clients who accessed host$_\textnormal{B}$'s web server have been infected with trojan$_\textnormal{Z}$'', ``trojan$_\textnormal{Z}$ has spread laterally in the network'').

\inlinedsectionitblock{Action Subspace} encodes the space of \emph{defense mechanisms} and \emph{protective measures} available to cyber defendants. Examples of defense mechanisms include proactive monitoring (e.g., update a sensor's policy), blocking (e.g., updating firewall rules), statically analysing deployed software for malware and vulnerabilities~\cite{jang:2012sp}, sandboxing~\cite{kirat:2014sec,kirat:2015ccs}, and deploying deceptive responses across the software stack~\cite{Araujo2014,Taylor2018,Zhang2020}. Protective measures include disseminating detection models and policies across endpoints, and sharing hunting outcomes with partner organizations.

\vspace{10pt}To reduce the impact of false positives, our framework enriches this decision model with threat intelligence and domain-specific information to weed out legitimate behaviors and identify new attack patterns. Upon verification, each malicious hypothesis elicits a protective measure harnessed from the action subspace.
Our implementation shows that this model is practical and can automate hypothesis generation for threat hunting tasks.

%% file: sections/hunting-model.tex
\section{Cyber Threat Hunting Model}
\label{sec:model}

\input{sections/hunting-model.definition}



%% file: sections/hunting-model.definition.tex
\newcommand\den[1]{{[\![ #1 ]\!]}}

\newcommand\cec{{c{\textnormal{\tiny\sf\&}}c}}
\newcommand\beac{\detector_{beacon}}
\newcommand\kge{\hunter_{KGE}}
\newcommand\impact{\hunter_{impact}}
\newcommand\ceccase{\hunter_{\cec}}
\newcommand\cecman{\manifold_{\cec}}
\newcommand\malwareman{\manifold_{malware}}
\newcommand\decoy{\manifold_{decoy}}
\newcommand\sand{\manifold_{sand}}
\newcommand\forensics{\verifier_{forensics}}
\newcommand\analytics{\verifier_{analytics}}

\newcommand\HTTP{\textnormal{HTTP\raise1.2pt\hbox{\scriptsize/}S}}
\newcommand\URL{\textnormal{ tokens}}
\newcommand\bins{\textnormal{\sc Bin}}
\newcommand\hfact[1]{\textnormal{\sc Intel$^{#1}$}}
\newcommand\malware{\textnormal{\sc Intel$^{malware}$}}
\newcommand\malinfo{\textnormal{\sc Info$^{malware}$}}
\newcommand\cecinfo{\textnormal{\sc Intel$^{\cec}$}}
\newcommand\cecmalinfo{\textnormal{\sc Intel$^{\cec, malware}$}}
\newcommand\cecip{{1.2.3.4}}

\newcommand\step[1]{{\mathrel{\xrightarrow[]{#1}}_{\mathrm{1}}}}
\newcommand{\append}{::}

\newcommand\huntstep[1]{{\vspace{6pt}\footnotesize #1 \vspace{6pt}}}

\newcommand*\conj[1]{\overline{#1}}


Our cyber threat hunting model (CTHM) conceptualizes  threat detection, analysis, and response as first-class components of the threat hunting process. In this model,
a \emph{hunt} is the tuple $(\kb, \hyps, \detectors, \hunters, \manifolds, \verifiers, \actions)$, where:

\begin{itemize}[itemsep=0pt]
\item $\kb=\rawdata\cup\intel$ is a set comprising \emph{internal data} $\rawdata$, sourced from monitoring and asset management infrastructures, and \emph{intelligence} data $\intel$, sourced from \emph{external threat intelligence}, \emph{domain knowledge}, or derived from hypotheses \emph{verified} during the threat hunting process.
\item $\hyps=\dhyps\cup\thyps$ is a set comprising detection ($\dhyps$) and threat ($\thyps$) hypotheses. A hypothesis $h\in\hyps$ can optionally define a confidence level, which can be derived from a detection event or predefined by the the threat analysis.
\item $\detectors$ is a set of \emph{detectors}, comprising threat detection analyses, $\detector\in\detectors, \detector:\kb\rightharpoonup\dhyps$
\item $\hunters$ is a set of \emph{case manifolds}, or hypothesis generators,  $\hunter\in\hunters, \hunter:\dhyps\times\kb\rightharpoonup\thyps$
\item $\manifolds$ is a set of \emph{decision manifolds}, $\manifold\in\manifolds, \manifold:\hyps\times\kb\rightharpoonup\actions$, where $\actions$ denotes the \emph{action subspace} of $\manifold$.
\item $\verifiers$ is a set of \emph{hypothesis verifiers} accepting or rejecting hypotheses, $\verifier\in\verifiers, \verifier:\hyps\times\kb\rightharpoonup\intel$
\end{itemize}

\noindent A \emph{case graph} $\mathcal{G}$ stores the state of a hunt, including the history of executed hunting steps and the outcomes produced during the execution of a hunt. Hunts and case graphs are two central concepts in our threat hunting model, analogous to programs and computation in traditional programming languages. Similarly, knowledge and hypotheses compare to abstract data types.

\contrib{We plan to explore this duality to enable powerful and expressive representations of  hunt workflows, affording threat hunters a suite of new tools---including a description language and its execution environment---to effectively unveil complex attack plots and help organizations protect their crown jewels. We plan to investigate deeper implications of the parallel between programming languages theory and threat hunting, such as leveraging type systems and enabling machine-checked proofs, or guarantees, of threat hunting predicates via application of the Curry-Howard correspondence~\cite{howard1980formulae}.}

A vector-valued objective \emph{cost function} $\mathcal{C}:\actions\rightarrow\mathbb{R}^{k}$, $\mathcal{C}(a)=(\mathcal{C}_{1}(a),\cdots,\mathcal{C}_{k}(a))$ maintains the cost values associated with the execution of each protective action $a\in\actions$, where $k\geq 1$ is the number of cost estimators, which can be derived from context-specific proxies (e.g., financial losses, liability risks) or computed as intrinsic functions of the action (e.g., employed resources, analysis time). Since $\mathcal{C}$ depends on the environment and can change during the course of a hunt, we write $\mathcal{G}\den{\mathcal{C}}$ to denote the cost function associated with a given hunt state. This choice of design for $\mathcal{C}$ explicitly models cost as a \emph{situation-specific} and \emph{user-specific} concept, and enables the application of \emph{normative} (i.e., prescriptive) theories to drive hypothesis generation in threat hunting scenarios.

\contrib{We plan to explore the applicability of decision theory and optimization to decision manifolds, and investigate minimal, Pareto-optimal sets of cost estimators for different application domains (e.g., enterprise, cloud, IoT).}


\inlinedsectionitblock{Manifold Selection.}
CTHM enforces that its different analysis components (detectors, manifolds, and verifiers) operate on well-defined abstractions (knowledge, hypotheses, actions) and interfaces. This ensures that hunts are always \emph{consistent} by construction---the model doesn't allow type mismatches. The choice of candidate manifolds and actions to be executed therefore reduces to the predefined semantics of each component, which may involve weighting uncertainty, costs estimations, and context associated with the state of the hunt.

%

\begin{figure}[t]
  \centering
  \includegraphics[width=1\columnwidth]{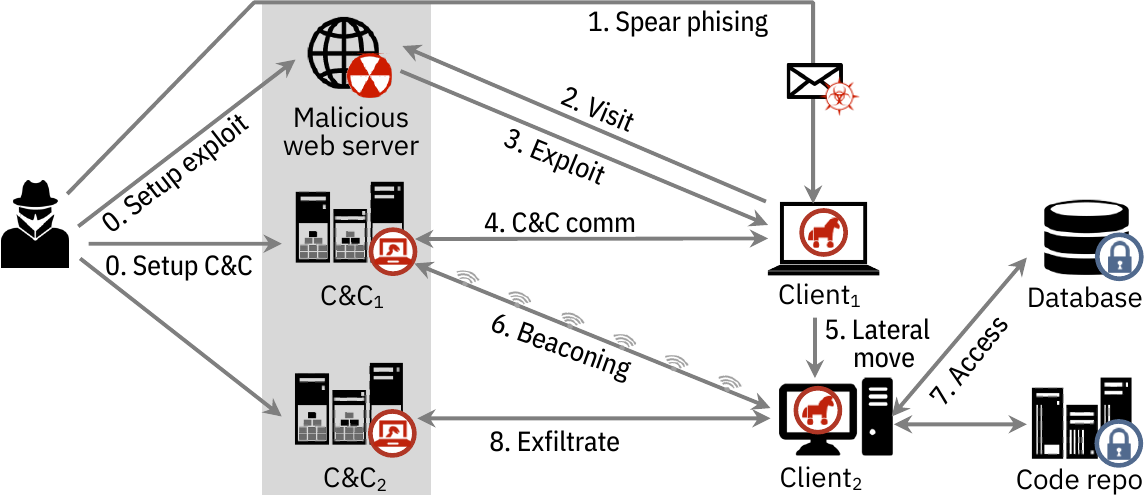}
  \caption{Attack scenario}
  \label{fig:attack-scenario}
\end{figure}

\inlinedsection{Attack Scenario.} \fig\ref{fig:attack-scenario} depicts an attack campaign, which comprises malware distribution servers and C\&C endpoints~(\circled{0}). The targeted attack starts with delivering a malicious web link via a spear phishing tactic~(\circled{1}). When the victim is tricked to follow the web link~(\circled{2}), the attacker's malicious web server exploits the browser's vulnerability to install malware~(\circled{3}). Once the malware infects the user's device, it establishes a communication channel with a C\&C server using DGAs and/or fast-fluxing techniques~(\circled{4}). The malware moves laterally to other devices to locate critical assets in the company network (e.g., by leveraging a zero-day exploit ~(\circled{5}). The malware also beacons to a C\&C server to report its presence and receive further instructions (\circled{6}), and conducts internal reconnaissance to identify and gain access to critical assets (e.g., code repositories) using stolen credentials~(\circled{7}). Finally, the malware exfiltrates sensitive information to an attacker-controlled remote server~(\circled{8}).


%% file: sections/case-study.tex
\section{Evidential Threat Hunting Semantics}
\label{s:case-study}

To illustrate these concepts, let
$(\kb, \hyps, \detectors, \hunters, \manifolds, \verifiers,
\actions)^{(0)}$ be the initial \emph{hunt state} of the threat hunting scenario depicted in \fig\ref{e:threatHuntingScenario}.
This example describes an attack campaign where an adversary has successfully compromised a small network of 10 client endpoints with a customized variant of Zeus~\cite{zeus}. Both network and endpoints are monitored for HTTP/S flows and system logs. Protective actions available to defenders include quarantining compromised endpoints to impede threat propagation across the network, containing and monitoring suspicious processes into a transparent sandbox, misdirecting attackers to decoys, actively seeking and hardening vulnerable targets in the network, and sharing threat information with partner organizations.

\input{sections/hunt}

\inlinedsectionit{Case Fragment 1: From beaconing to malware detection}

\noindent The hunt begins when  $\beac$ observes beaconing behavior between remote host \cecip{} and client$_{1}$, resulting in a new detection hypothesis~(step~\circled{0}--\circled{1}).

\huntstep{
$
(\kb, \hyps, \detectors, \hunters, \manifolds, \verifiers, \actions)^{(0)}
\step{\beac}\\
(\kb, \hyps^{\textit{\tiny{i}}}\mapsto\hyps\append{beacon(\cecip, \textnormal{client}_{1})}, \detectors, \hunters, \manifolds, \verifiers, \actions)^{(1)}
$
}

\noindent A case manifold $\kge$ explores a cognitive knowledge graph to generate two threat hypotheses, which predicate that (a)~host \cecip{} is a C\amper{}C server (step~\circled{1}--\circled{2}), and that (2)~client$_{1}$ is infected with Zeus~(step~\circled{2}--\circled{3}).

\huntstep{
$
(\kb, \hyps^{\textit{\tiny{i}}}, \detectors, \hunters, \manifolds, \verifiers, \actions)^{(1)}
\step{\kge}\\
(\kb, \hyps^{\textit{\tiny{ii}}}\mapsto\hyps^{\textit{\tiny{i}}}\append{\cec(\cecip)}, \detectors, \hunters, \manifolds, \verifiers, \actions)^{(2)}
$\\

$
\hspace{10.3em}
\step{\kge}\\
(\kb, \hyps^{\textit{\tiny{iii}}}\mapsto\hyps^{\textit{\tiny{ii}}}\append{infected(\textnormal{client}_{1}, \textnormal{zeus})}, \detectors, \hunters, \manifolds, \verifiers, \actions)^{(3)}
$
}

\noindent To verify these hypotheses, the threat hunter leverages an automated security analytics tool $\analytics$ (e.g., a Security Knowledge Graph) to assert that \cecip{} indeed belongs to a known C\amper{}C server (step~\circled{3}--\circled{4}). Meanwhile, an endpoint forensics analysis $\forensics$ reveals that client$_{1}$ has been infected by a specific self-propagating variant of Zeus (step~\circled{4}--\circled{5}).
The analyst accepts these threat hypotheses, which become new knowledge.

\huntstep{
$
(\cdots)^{(3)}
\step{\analytics}
(\kb^{\textit{\tiny{i}}}\mapsto\kb\append{\cec(\cecip)},\\ \hyps^{\textit{\tiny{iv}}}\mapsto\hyps^{\textit{\tiny{iii}}}/{\cec(\cecip)}, \cdots)^{(4)}
$\\

$
\hspace{3.2em}
\step{\forensics}
(\kb^{\textit{\tiny{ii}}}\mapsto\kb^{\textit{\tiny{i}}}\append{infected(\textnormal{client}_{1}, \textnormal{zeus})},\\ \hyps^{\textit{\tiny{v}}}\mapsto\hyps^{\textit{\tiny{iv}}}/{infected(\textnormal{client}_{1}, \textnormal{zeus})}, \cdots)^{(5)}
$
}

\inlinedsectionit{Case Fragment 2: Lateral movement assessment}

\noindent To assess the scope and impact of this attack campaign, a case manifold $\impact$ hypothesizes whether the trojan spreads itself laterally to other clients in the network (step~\circled{5}--\circled{6}). Using information acquired in prior hunts about this specific variant of Zeus, $\impact$ gathers a list of all the clients whose system logs reveal network file system access from client$_{1}$ (e.g., SMB), and performs a forensics analysis on each client to confirm the propagation of the trojan. The outcome of this analysis reveals that while client$_{2}$ has been compromised, no evidence of this particular variant of Zeus was found in client$_{7}$ (step~\circled{6}--\circled{7}).

\huntstep{
$
(\cdots)^{(5)}
\step{\impact}
(\kb^{\textit{\tiny{ii}}}, \hyps^{\textit{\tiny{vi}}}\mapsto\hyps^{\textit{\tiny{v}}}\cup\{infected(\textnormal{client}_{2}, \textnormal{zeus}),\\ infected(\textnormal{client}_{7}, \textnormal{zeus})\}, \cdots)^{(6)}
$\\

$
\hspace{3.2em}
\step{\forensics}
(\kb^{\textit{\tiny{iii}}}\mapsto\kb^{\textit{\tiny{ii}}}\append{infected(\textnormal{client}_{2}, \textnormal{zeus})},\\ \hyps^{\textit{\tiny{vii}}}\mapsto\hyps^{\textit{\tiny{vi}}}/{infected(\textnormal{client}_{7}, \textnormal{zeus})}, \cdots)^{(7)}
$
}




\input{tables/valuations}

\inlinedsectionit{Case Fragment 3: Deployment of protective measures}

\noindent Following this initial assessment, action manifolds subscribing to the new facts generated in prior hunt steps start a deliberation process to decide on the next hunt step.
In this case, the optimality criteria for the decision combine threat hunting \emph{goals} set beforehand by the threat hunter, \emph{constraints} imposed by the current hunt state $\mathcal{G}^{(7)}$, and \emph{cost} and \emph{risk} estimations derived from a simple, yet relevant cost model. For explanatory precision, let
$\mathcal{G}^{(7)}\den{\mathcal{C}}=(\conj{\mathcal{C}_{1}\cdots\mathcal{C}_{6}})$ denote the cost function associated with the hunt, where:

{
\vspace{6pt}
\footnotesize
$\mathcal{C}_{k}:\mathcal{A}\rightarrow{\{low,moderate,high\}}$, \\ $k=1\cdots6$, and
$\left\{
\begin{tabular}{@{}l@{}}
    $\mathcal{C}_{1}:$ system downtime \\ 
    $\mathcal{C}_{2}:$ allocated resources \\
    $\mathcal{C}_{3}:$ analysis time \\
    $\mathcal{C}_{4}:$ defender risk \\
    $\mathcal{C}_{5}:$ threat intel acquisition \\
    $\mathcal{C}_{6}:$ attacker risk \\ 
\end{tabular}
\right\}$
cost estimators.
\vspace{6pt}
}

\noindent Risk is encoded as cost estimations, and attack cost estimations output negative costs. For simplicity, cost valuation results fall into one of three categorical values: $low$, $moderate$, or $high$.

\tab\ref{tab:costmodel} summarizes $\mathcal{G}^{(7)}\den{\mathcal{C}}$'s cost valuations and the manifold deliberation process.
Leveraging this analysis, action manifold $\malwareman$ performs its deliberation over infected endpoints client$_{1}$ (a critical asset) and client$_{2}$ (a crown jewel), resulting in the three prescriptive actions to the risk-averse threat hunter (step~\circled{7}--\circled{8}). Finally, $\cecman$ shares the newly uncovered attack campaign information $\{\cec~hunt\}$ with its organization's partners (step~\circled{8}--\circled{9}).



\huntstep{
$
(\cdots)^{(7)}
\step{\malwareman}
(\kb^{\textit{\tiny{iii}}},
\hyps^{\textit{\tiny{vii}}},\cdots,\{\textsc{contain}[client_{1}], \\  \textsc{quarantine}[client_{2}],\textsc{fortify}[decoy_{1\textnormal{--}25}]\})^{(8)}
$\\

$
\hspace{3.2em}
\step{\cecman}
(\kb^{\textit{\tiny{iii}}}, \hyps^{\textit{\tiny{vii}}},\cdots,\{\textsc{share}[\hspace{0.1em}\{\cec~hunt\}\hspace{0.1em}]\})^{(9)}
$
}
\vspace{-6pt}

%% file: sections/hunt.tex
\begin{figure}[t]
{
\scriptsize
\begin{mdframed}[style=frame,innerleftmargin=0pt,innerrightmargin=-10pt]
\begin{align*}
\textit{Internal\quad}			\rawdata 	&=  \{endpoints: \{\textnormal{clients$_{1-10}$}\},&\\
                 	 				&			monitoring: \{\textnormal{\HTTP~records, syslogs}\}\} &\\
\textit{Intelligence\quad}			\intel		&=	\{{c{\textnormal{\footnotesize\sf\&}}c}: \{\cecip\},&\\
					  						&	malware: \{(\textnormal{zeus}, \textnormal{014e7...7bbb})\}\} &\\
\textit{Hypotheses\quad}			\hyps		&= 	\bot \\
\textit{Detectors\quad} 			\detectors	&=	\{\beac:~\textnormal{\HTTP}\rightharpoonup\dhyps^{beacon}\} &\\
\textit{Cases\quad}   		\hunters	&=	\{\kge:~\dhyps^{beacon}\times\kb\rightharpoonup\thyps^{\cec, malware}, &\\
												&	\hspace{1.9em} \impact:~\underline{\hspace{1em}}\times\kb\rightharpoonup\thyps^{\cec, malware}\} &\\
\textit{Decisions\quad}		\manifolds	&=	\{\cecman:~\thyps^{\cec}\times\kb\rightharpoonup\actions, &\\
								   				&   \hspace{1.9em}\malwareman:~\thyps^{malware}\times\kb\rightharpoonup\actions\} &\\	
\textit{Verifiers\quad} 			\verifiers	&=	\{\forensics:~\thyps^{malware}\times\kb\rightharpoonup&\\
& \hspace{10em}\{\thyps^{malware}_{accepted}, \thyps^{malware}_{rejected}\}, &\\
												&   \hspace{1.9em} \analytics:~\thyps^{\cec}\times\kb\rightharpoonup&\\
												& \hspace{10em} \{\thyps^{\cec}_{accepted},\thyps^{\cec}_{rejected}\} &\\
Actions\quad 						\actions	&=	\{\textsc{quarantine, contain, misdirect, fortify, share}\} &
\end{align*}
\end{mdframed}
\vspace{-10pt}
}
\caption{An initial hunt state for a threat hunting scenario\label{e:threatHuntingScenario}}
\vspace{-18pt}
\end{figure}

%% file: tables/valuations.tex
\newcommand\cost[1]{$\mathcal{C}_{#1}$}
\newcommand\costzero{\Circle}
\newcommand\costlow{\LEFTcircle}
\newcommand\costhigh{\CIRCLE}
\newcommand\costnzero{{\color{red}\Circle}}
\newcommand\costnlow{{\color{red}\LEFTcircle}}
\newcommand\costnhigh{{\color{red}\CIRCLE}}

\begin{table*}[t]
\centering
\footnotesize
\caption{Cost valuations for $\mathcal{G}^{(7)}\den{\mathcal{C}}$ and action deliberation considerations}
\begin{tabular*}{0.95\textwidth}{ cl@{\hskip 1cm}lllllll }
  \toprule
  \textbf{Type} & \textbf{Action} & \cost{1} & \cost{2} & \cost{3} & \cost{4} & \cost{5} & \cost{6} & \textbf{Deliberation} \\ 
  \midrule
  \multirow{2}{*}{\parbox[t]{2cm}{\centering threat\\ mitigation}} 	& \textsc{quarantine} & \costhigh & \costzero & \costzero & \costlow &  \costnzero & \costnzero & if target is a crown jewel \\[0.2em]
																	& \textsc{contain} & \costzero & \costzero & \costlow & \costlow & \costnlow &  \costnlow & if target can't tolerate downtime\\ 
  \midrule
  \multirow{2}{*}{\parbox[t]{2cm}{\centering proactive\\ hunting}} 	& \textsc{misdirect} & \costzero & \costzero & \costzero & \costhigh & \costnzero & \costnhigh & if defender is resource-constrained \\[0.2em]
																	& \textsc{fortify} & \costzero & \costhigh & \costzero & \costzero & \costnlow & \costnlow & if defender is averse to risk \\ 
  \midrule
   intelligence														& \textsc{share} & \costzero & \costlow & \costzero & \costzero & \costnlow & \costnzero & to inform partners \\ 
  \bottomrule
  \multicolumn{9}{r}{{\scriptsize \textsc{key:} \costzero=low \quad \costlow=moderate \quad \costhigh=high \quad (values in {\color{red}red} represent attacker cost)}} \\
\end{tabular*}
\label{tab:costmodel}
\end{table*}

%% file: sections/implementation.tex
\section{Implementation}
\label{sec:thdl}
\input{sections/hunting-model.language}

\label{sec:hunting-computation}
\input{sections/hunting-model.computation}

%% file: sections/hunting-model.language.tex
\inlinedsection[0pt]{Threat Hunting Description Language.}
We implemented a domain-specific language~\cite{Shu2018} and data model~\cite{Taylor2020} to facilitate the creation and execution of complex threat hunting flows. 
This Threat Hunting Description Language (THDL) 
allows security analysts to uniformly \textit{i)}~express domain knowledge, \textit{ii)}~retrieve observations for further threat hypothesis development, \textit{iii)}~verify hypotheses against the data, and \textit{iv)}~execute decision manifolds (\s\ref{sec:model}). To support sharing and reuse of hunt flows, the language satisfies the following design goals:

\begin{itemize}[itemsep=0pt]

	\item \textit{Composability}: larger, more complex hunts with a compounded effects of decision manifolds can be created by composition of smaller, simpler hunting steps and decision manifolds.



	\item \textit{Explainability}: each hunt step as well as a collection (code block) of steps are human-understandable and support rich visual representations to facilitate quick assimilation of hunt workflows by novice threat hunters.
	



	\item \textit{Partial updatability}: the system  minimizes re-execution and caches partial results when parts of the composed hunt flows is changed and needs to be re-evaluated.


\end{itemize}

%% file: sections/hunting-model.computation.tex
\inlinedsection{Automated Reasoning.}
Our framework also implements a language runtime and hunt execution 
pipeline that enables
efficient and scalable hunt execution through
automated reasoning and case prioritization (\s\ref{sec:model}). Hunts and case graphs form the basis of a threat hunting computation model that defines \emph{hunt programs} orchestrated by the framework. For each case initiated by
detection hypotheses $\dhyps\subseteq\hyps$, a corresponding hunt program $(\kb, \hyps, \detectors, \hunters, \manifolds, \verifiers, \actions)$ is instantiated. 
The framework embodies three main components:
\begin{itemize}[itemsep=0pt]

	\item \textit{Semantic steps}: The execution of the {hunt program} emits a sequence of operational steps called \emph{semantic steps} (\s\ref{s:case-study}). A {semantic step} represents a unit of reasoning task in a hunt analysis. These steps may involve simple case manifold executions, such as retrieving additional data from different data sources, or performing complex analytics or running other existing hunt programs.

	\item \textit{Semantic reasoning}: To reason about the current case graph and properly prioritize the execution of the next {semantic step} out of many potential steps, our implementation performs a signal propagation analysis called \emph{signal flow}. Signal flow propagates multiple signals in the case graph representing different semantic meanings, such as relevance and toxicity (badness). The flow of a signal is influenced by various factors, such as behavioral, temporal, and veracity aspects, resulting an annotated case graph with semantically-relevant hunt paths. By combining signal flow with a \emph{cost and reward} strategy, our framework produces a dynamic and adaptive sequence of steps guided by the reasoning semantics. 

	\item \textit{Hypothesis generation}: The framework distills competing hypotheses, chains hypotheses from multiple case manifolds, and ranks and summarizes them before reporting to a human analyst. 

\end{itemize}

%% file: sections/protect-measures.tex
\inlinedsection{Protective Responses.}
Automated threat responses generated by our workflows (\s\ref{s:case-study}) must not allow any proactive measure to unwittingly be used to attack the network it is trying to protect. Moreover, such measures must try to identify
all network hosts involved in an attack campaign and quarantine them at once so that no backdoors remain open for the attacker. To satisfy these constraints, our framework augments traditional threat responses with deceptive capabilities to anticipate and foil attempted attacks.  These deceptive responses include dynamically sandboxing or terminating attacker sessions~\cite{Araujo2020}, implementing filesystem separation~\cite{Araujo2018,Taylor2018}, altering the network topology~\cite{Stoecklin2018}, and embedding decoys into software applications~\cite{Araujo2014,araujo19acsac,Araujo2021}. Overall, our proactive measures fall into five categories: \emph{containment} (e.g., sandboxing), \emph{misdirection} (e.g., deception), \emph{quarantine} (e.g., isolation), \emph{fortification} (e.g., vulnerability scanning), and \emph{sharing} (e.g., threat reports).

%% file: sections/conclusion.tex
\section{Conclusion}
\label{sec:conc}
This paper conceptualized and outlined the implementation of an evidential   multi-criteria cyber reasoning framework for threat hunting. We are currently evaluating this framework with diverse attack scenarios and data to optimize threat hunting workflows in a large security operations center.